# An IAC Approach for Detecting Profile Cloning in Online Social Networks

MortezaYousefi Kharaji[1] and FatemehSalehi Rizi[2]

[1]Deptartment of Computer and Information Technology Engineering, Mazandaran University of Science and Technology, Babol, Iran
[2]Department of Computer Engineering and Information Technology, Sheikhbahaee University of Isfahan, Isfahan, Iran

## Abstract

*Nowadays, Online Social Networks (OSNs) are popular websites on the internet, which millions of users register on and share their own personal information with others. Privacy threats and disclosing personal information are the most important concerns of OSNs' users. Recently, a new attack which is named Identity Cloned Attack is detected on OSNs. In this attack the attacker tries to make a fake identity of a real user in order to access to private information of the users' friends which they do not publish on the public profiles. In today OSNs, there are some verification services, but they are not active services and they are useful for users who are familiar with online identity issues. In this paper, Identity cloned attacks are explained in more details and a new and precise method to detect profile cloning in online social networks is proposed. In this method, first, the social network is shown in a form of graph, then, according to similarities among users, this graph is divided into smaller communities. Afterwards, all of the similar profiles to the real profile are gathered (from the same community), then strength of relationship (among all selected profiles and the real profile) is calculated, and those which have the less strength of relationship will be verified by mutual friend system. In this study, in order to evaluate the effectiveness of proposed method, all steps are applied on a dataset of Facebook, and finally this work is compared with two previous works by applying them on the dataset.*

## 1.Introduction

Social network websites are defined as web services that allow users to make public and semi-public profiles in a bounded system, to build a list of users with whom have a kind of common relationship, and to search in their friends' lists [1]. One of the most important challenges of observing friends' information is threatening users' security and privacy. An adversary can cause many problems by exploiting users' information. This data may contain users' financial information which adversary can use them to do identity theft attacks, or may contain users' medical background such as healthy status, diagnosis or treatment records [2].

Recently, a new kind of attack which is named Identity Clone Attack is detected on OSNs that makes fake identities of specific users. The basic goals of the adversary in this attack are obtaining victim's friends' personal information by forging real user profile, and increasing trust among mutual friends to do more defrauding in the future [3]. Two kinds of these attacks are already defined: first one is Single-Site Profile Cloning, and the next one is Cross-Site Profile Cloning. In the first attack, adversary forges the real user profile in the same social network and use this cloned profile to send friend request to users' friends. An unaware user may think this

DOI : 10.5121/ijnsa.2014.6107    75

request is from a familiar user hence she/he will confirm it and his/her personal information will be accessible for adversary. The next attack is cross-site profile cloning, as it shown in Figure1, the adversary detects a user with his/her friends in network A, then make a clone profile with his/her attributes in network B which user has not made account yet. The adversary sends friend requests to the victim's friends in network B. Victim's friends think they know the sender of requests and confirm them, and as soon as they confirm the request, the adversary will thieve their personal information. The adversary uses this information to make other clone profiles or to deceive others in the future. Detecting this kind of attack is very difficult for service providers and profiles owners, because service providers think it is a new user which is registering in these websites [4]. Discovering cloned profiles with more precise methods can bring more security for users who are using social networks, and also cause an increasing movement for service providers to improve their security level in the services they provide on their platforms [5].

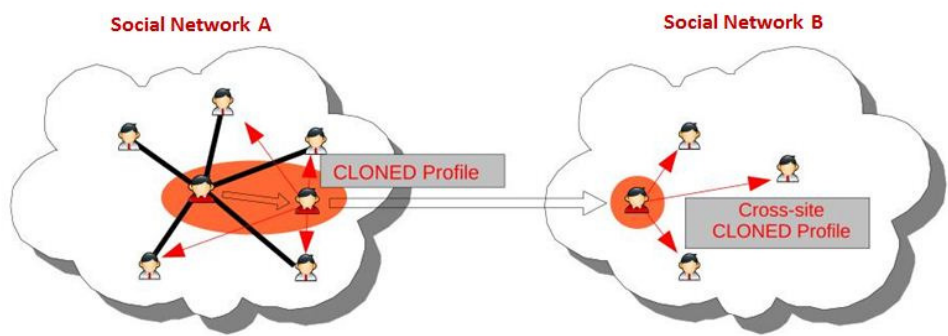

Figure 1. Single-site profile cloning and cross-site profile cloning attacks [5]

The rest of the paper is organized as follows: in section 2,a short review on related works are expressed and section 3 presents the proposed method for detecting cloned profiles in complete details. In section 4, to evaluate the applicability of proposed method, it is applied on a dataset of Facebook and it is also compared with previous works in section 5. Finally, in section 6, the paper is concluded and some feasible future works are discussed.

## 2.Related Works

Many social networks have a weak user to user authentication mechanism that are mostly based on presented information such as name, photos, and a set of social links. This causes the misuse of profile cloning attack to make fake social links. Bhumiratana in [6] presented a model to exploit of available weak trust in social networks. This model saves the authority of an online fake identity which made by profile cloning attack to obtain more personal information. This research proposed an attack methodology to use cloned profiles and to do reliable interactions among selected users. Proposed model uses an array of attacking techniques to make a permanent and automatic cloned identity of real users on social networks so that are able to get personal data in a specific period of time. This proposed system works among different social networks.

Jin et al. in [7] proposed an active detection framework to detect cloned profiles.An intelligent fake identity not only forges users' attributes, but may add victim's friends into his friend network too. According to similarity of attributes and users' friend list there are two ways for defining similarity measure among real identity and fake identities. One of them is basic profile similarity and the next one is multiple-faked identities profile similarity. In this research, according to the similarity of profiles, a framework for detecting cloned profiles on social network is proposed which contains of three steps: first step is to search and separate identities as a set of profiles, as the entry of search is a profile attributes. Second step is detecting suspicious profiles by using profile similarity schemas, and third step is deleting cloned profiles from friend list. In detecting process adjusting a set of parameters can help to do a correct detection in different social networks.

Kontaxis et al. in [8] offered a tool which is able to automatically search and detect cloned profiles in OSNs. The concept key of their approach is using user-specific data which is extracted from real user profile in social network. In this approach, finally a list of profiles which are probably cloned with similarity scores is presented to user. A string matching algorithm is used to define the similarity of attributes between two profiles and assign similarity score for each candidate identity. In this method detecting cloned profile contains three steps as follows: information Distiller, profile hunter, and profile verifier.

Gani et al. in [9] discussed a piece of work which intends to provide some insights regarding the resolution of the hard problem of multiple identities detection. Based on hypothesis that each person is unique and identifiable whether in its writing style or social behavior, they proposed a framework relying on machine learning models and a deep analysis of social interactions, towards such detection.

Most of the current research has focused on protecting theprivacy of an existing online profile in a given OSN. Instead, Conti et al. in [10]noted that there is a risk of not having a profile in the last fancysocial network. The risk is due to the fact that an adversary maycreate a fake profile to impersonate a real person on the OSN.The fake profile could be exploited to build online relationshipwith the friends of victim of identity theft, with the final target ofstealing personal information of the victim, via interacting onlinewith the friends of the victim.

## 3. The proposed approach

The detection approach is organized in 6 steps as follows:

### 3.1. Discovering community the social network graph

In many social networking sites, network topological structure and attributes values are the complete information. Nodes represent users and edges represent the relationship among them. In each node, there are some attributes such as name, gender, education, interests, location and social activities. It is obvious that network topological structure and attribute information can be used to identify some hidden patterns in communities. In this study, IAC clustering algorithm [11] is applied to detect communities in social network graphs. Figure 2 shows a pseudo code of the algorithm where it accepts an attribute augmented graph and return a clustered graph as output.

1. **Input:** $G, \alpha$
2. **Output:** clusters
3. $A \leftarrow \text{Adj}(G)$
4. $K = \alpha \times |E(G)|$
5. Compute the attributes similarity matrix, $C$
6. $S_{ij} = 1$ if $(i,j) \in TopKpair(C)$, 0 otherwise
7. $W \leftarrow A + S$
8. clusters $\leftarrow$ Apply Markov Clustering on $W$
9. **Return** clusters

Figure 2. IAC Clustering Algorithm [11]

International Journal of Network Security & Its Applications (IJNSA), Vol.6, No.1, January 2014An augmented graph is a graph G = (V, E, χ), where V = {$v_1$, $v_2$, $v_3$, …,$v_n$ } is the set of nodes and n = |V| denotes the number of nodes in the graph, E ⊂ V× V is the set of edges,E = {($v_i$, $v_j$): $v_i$, $v_j$∈ V}, and χ∈ $R^{|v| \times d}$is the nodes attribute matrix. First of all, the algorithm creates the similarity matrix C, then according to K (K = α× E) it adds the set of edges to the graph and the elements which belong to these edges are set to 1 in matrix S. As well as matrix W is made by summation of S and A. To this end, a weighted graph is clustered by MCL algorithm that is demonstrated in Figure 3. MCL is a clustering algorithm [12] based on stochastic flows on the graph and in order to execute it, first, transition matrix should be made from weighted graph obtained through matrix W. This algorithm includes expansion and inflationoperations on stochastic matrixes such that the expansion is calculated as M×M and the inflation increases the M's elements to amount of r (r > 1), then normalizes each column. Eq. 1 indicates how the inflation operation works, after normalizing the summation of each column will be 1.

$$(\Gamma_r M_{pq}) = \frac{(M_{pq})^r}{\sum_{t=1}^{k}(M_{pq})^r} \qquad (1)$$

MCL is started from a standard flow matrix and the two operations apply it alternatively until the output matrix gets a stable state and it will not be changed when the operations are applied again. After,allof clusters are determined in the rows of the stable matrix.

```
Algorithm 1 MCL
A := A + I   // Add self-loops to the graph
M := AD^{-1} // Initialize M as the canonical transition matrix
repeat
    M := M_{exp} := Expand(M)
    M := M_{inf} := Inflate(M, r)
    M := Prune(M)
until M converges

Interpret M as a clustering
```

Figure 3. MCL clustering algorithm [12]

### 3.2. Extraction user's attribute

In this stage, the user's information is extracted from his/her legitimate profile in online social network. At the start, the user's profile is analyzed then it is specified that which parts of user's profile can be regard as user-specific. This information is used to construct queries in search engines of social networks. The extracted information is includes name, gender, location, education, email and etc. social networks owner and service provider have complete access to users' data and can exploit user-specific from her/his profile easily.

### 3.3. Search in community

In step1, the socialgraphwasclusteredconcerning to users' attribute similarities. In this stage for finding similar profiles to real user's profile, the cluster which is belong to real user is marked then all of similar profiles are searched by name attribute. The search result is the list of profiles with similar or same name to real profile.

78

## 3.4. Selecting profile

In this stage, the profiles which have mutual friends with victim (real profile) are picked up among founded profiles in step 3. Mutual friends are the friends who exist in the victim's friend list and in the friend list of each candidate profile in the same time. Since, in profile cloning attacks many friend requests are sent to victim's friends, it is obvious they have some common friends with victim [4]. Hence, only profiles which have mutual friends with victim are chosen for continuing next steps.

## 3.5. Computing strength of relationship

In step 5, all of nodes' edges which was acceded in this stage, are weighted considering to the number of common active friends, shared Urls and page-likes among users. Formally, the social network can be defined as a weighted graph G = (V, E, W), where V is the set of profiles, E ⊆ V × V is the set of edges, and W ⊆ $\Re$ is a set of weights are assigned to edges. For each node v ∈ V, a 3-dimentional feature vector is defined as it is included in the number of active friends, page likes and common shared URLs. Therefore, weight of each edge $e_{ij} = (v_i, v_j)$ is calculated as summation of common actives friends, page likes and common shared URLs between nodes $v_i$ and $v_j$. Further details presented how the weights can compute come in the following parts [13].

3.5.1 Active friends:

This measure takes the interaction frequency of a user with his/her friends in the network. For a user $V_i$ with $F_i$ as the set of friends, the set of active friends $F_i^a$ can be computed as an interaction between the set $F_i$ and the set of friends of $V_i$ who were either contacted by $V_i$ or those who interacted with $V_i$ through wall posts, comments or tags. It can be defined using Eq. 2 in where $I_i$ is the set of users with whom $V_i$ has interactions in the network. For a node $V_i$ the value of the "active friends" feature is taken as the cardinality of the set of its active friends $F_i^a$. Similarly, the set of common active friends in the network with whom a pair of users $v_i$ and $v_j$ have interacted is calculated as the intersection of their active friends $F_i^a$ and $F_j^a$, respectively, as given in Eq. 3. For an edge $e_{ij} = (v_i, v_j)$, the value of the "active friends" feature is taken as the cardinality of the set of common active friends $F_{ij}^a$ [13].

$$F_i^a = F_i \cap I_i \quad (2)$$

$$F_{ij}^a = F_i^a \cap F_j^a \quad (3)$$

3.5.2 Pages-likes:

This feature computes the page likes frequency of the users in social network. For an edge $e_{ij} = (v_i, v_j)$, the common page likes of $v_i$ and $v_j$, $P_{ij}$, is calculated as the interaction of the sets of page likes of $v_i$ and $v_j$, as given in Eq. 4, and the page likes attribute value is calculated as the cardinality of the set $P_{ij}$ [13].

$$P_{ij} = P_i \cap P_j \quad (4)$$

3.5.3 URLs:

this feature captures the URL sharing patterns of the social networks users. For an edge $e_{ij} = (v_i, v_j)$, the common URLs of $v_i$ and $v_j$, $U_{ij}$, is calculated as the intersection of the set of URLs shared by $v_i$ and $v_j$. The URLs attribute value is calculated as a fraction of URLs commonly shared by them using Eq. 5 [13].

$$U_{ij} = \frac{U_i \cap U_j}{U_i \cup U_j} \quad (5)$$

On the basis of the above mentioned features, each edge $e_{ij} = (v_i, v_j)$, is assigned a weight $w(e_{ij})$ that is calculated as an summation of the individual feature value as given in Eq. 6. ‖represents the cardinality of the set [13].

$$w(e_{ij}) = |F_{ij}^a| + |P_{ij}| + |U_{ij}| \qquad (6)$$

Afterward, the weights are assigned to each edge in social network graph and strength of relationship is calculated between two nodes as follows:

3.5.4 Definition 1 (Friendship Graph) [14]

Given a social network G and a node v ∈ G.N, the friendship graph of v, denoted as FG (v), is a sub-graph of G where: (1) FG(v).N = {v} ∪ {n ∈ G.N | n ≠ v, ∃ e ∈ G.E, e = <v, n> }; (2) FG(v).E = {e = <v, n>∈ G.E | n ∈ FG (v).N} ∪ {e = <n, n′>∈ G.E | n, n′ ∈ FG(v).N }

3.5.5 Definition 2 (Mutual Friends Graph) [14]

Given a social network G and two nodes v, c ∈ G.N, the mutual friends Graph of v and c, denoted as MFG(v, c), is a sub-graph of G where: (1) MFG(v, c).N = {v, c} ∪ {n ∈ G.N | n ≠ v, n ≠ c, ∃ e, e′∈ G.E, e = <v, n>∧ e = <n, c> }; (2) MFG(v, c).E = {e = n, n′ ∈ G.E | n, n′ ∈ MFG (v, c).N}

For instance, Friendship graph of node 7 and mutual friends graph of 7 and 12 are shown in Figure 4.

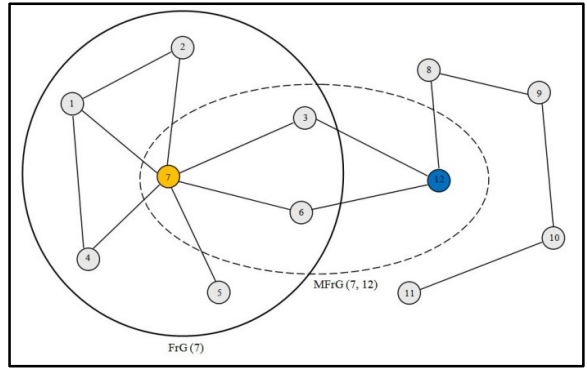

Figure 4. Friends and mutual friends graphs

3.5.6 Definition 3 (Strength of relationship between two nodes)

Given a social network G and two nodes v, c ∈ G.N, Let T = {MFG(v, c).E }, R = { FG(v).E }, P = {FG (c).E }. Strength of relationship between v and c is defined in Eq.7 as follows:

$$SR(v, c) = \frac{\sum_{r \in T} w_r}{\sum_{r_1 \in R} w_{r_1} + \sum_{r_2 \in p} w_{r_2}} \qquad (7)$$

Strength of relationship (SR)measure is calculated between each suspiciousprofile which hasmutual friends with victim. Inasmuch as an expert adversary attempts to make less suspicious by making social relationship and interactions with victim's friends. Strength of relationship measure is used to detect cloned identities because the real identities make more deep social activities than them as they mostly know each other in real life. They might get intimacy through relationships in real life or voice and video chat on the Internet for a while [15]. Therefore, real

International Journal of Network Security & Its Applications (IJNSA), Vol.6, No.1, January 2014

users contribute in social activities like commenting, sending message and tagging more than fakes and clearly they have higher SR comparing to cloned profiles. In the rest of this stage, nodes are sorted in a list by amount of SR as RS (v, $c_1$) < RS (v, $c_2$)<RS (v, $c_3$) <…<RS (v, $c_n$) and n is the number of profiles which have reached in step 5. Among these profiles, $c_1$ has the least SR and it will be sent to next step for verifying. If it does not identify as a cloned identity the next one in the list, $c_2$ will be gone to stage 6. This trend will be continued until the last profile in the list.

### 3.6. Decision making

Heretofore, some methods were presented to verify the suspicious identity in online social networks. In a primary approach, the ID number is asked from users for verification process. For example Identity Badge wants users to enter their passport number [16]. The social verification approach is presented by Schechter et al. [17] want users to design some questions to verify their friends and if a user answers most questions correctly he/she will be marked as a valid user. A proposed approach is verifying suspicious identities by mutual friends as it is wanted mutual friends to design some question concerning to background knowledge that they have obtained during their relationships. As well as these questions can be design by some social engineering teachings. It is evident that a cloned identity cannot answer the question correctly, specially the questions which are designed considering to users' background. Also similar identities (are not fake) send their own answers which are understandable for mutual friends as they come from real identities. Eventually, fake identities are identified and they remove or closed temporary by service provider as well as their friends receive some notification for existence a fake identity in their friend list. Figure 5demonstrates a view of proposed verification system.

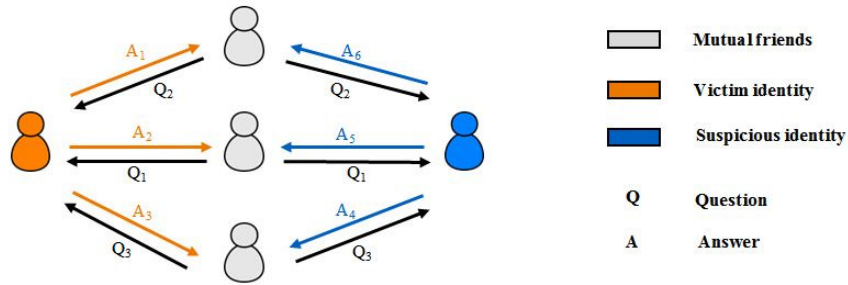

Figure 5. Verification system by mutual friends

The diagram of detection approach is shown in Figure 6.

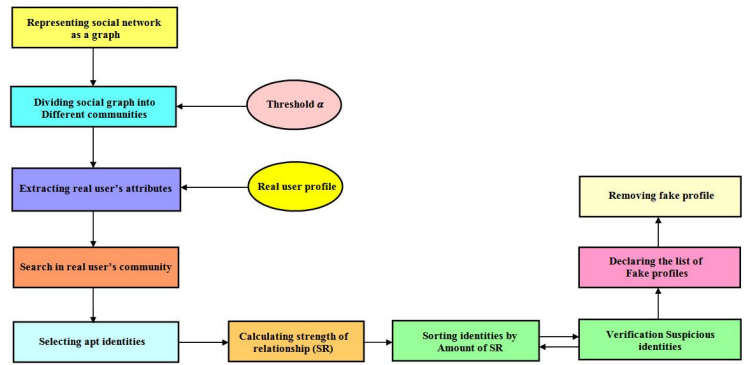

Figure 6. IAC Detection Approach

81

## 4. Experimental Results

In order to evaluate the proposed approach, an office dataset of Facebook users [18] is used and it is updated by adding user's attributes, shared Urls and page-likes. Verification the proposed approach is not possible for a normal social network user because only service providers haveaccess to users' original information and social network graph.Also some social networks have restrictions thus normal users cannot make clone profile easily [4]. There are 63,731 users in this dataset and 1,634,115 links among them thus each user has 25.6 relationship links on average. To evaluate the approach, it is assumed that there are some fake identities in this dataset and it is necessary to add themto dataset as victims. For demonstrating the detail of effectiveness of proposed approach, 20 users are selected from the dataset as their social graph is shown in Figure7 and their attributes are exhibited in Table 1.

Table 1. 20 users' attributes selected from dataset

| ID | Name | Gender | Education | | Work | | Birthday | Location | Relationship |
|---|---|---|---|---|---|---|---|---|---|
| | | | School | Degree | Employer | Position | | | |
| 32 | NikoParda | Female | Harvard University | PhD | East Man | Manager | 1979 | USA | Single |
| 35 | Sara Abraham | Female | Arcadia University | Master's | Owens | Web Developer | 1980 | USA | Single |
| 36 | Sara Abraha | Female | Carolina University | Master's | Owens | Web Developer | 1980 | USA | Single |
| 174 | David Ernox | Male | Michigan University | Master's | Qpass | Java Developer | 1984 | USA | Single |
| 463 | Sara Abram | Female | Michigan University | Master's | AppNet | Web Developer | 1985 | USA | Single |
| 1236 | Tom Banho | Male | Acaedia University | Bachelor | Xing | Network Manager | 1979 | USA | Married |
| 2411 | Rose Milan | Female | Koln University | PhD | Axvert | Manager | 1972 | USA | Single |
| 33 | Hanrry Dabuo | Male | Dublin High school | Diploma | Sonic | Secretary | 1970 | UK | Married |
| 34 | Rosa Morada | Female | Franklin High school | Diploma | Sonic | Bookkeeping | 1974 | UK | Married |
| 163 | Charls Selvin | Male | Pietersburg University | Bachelor | Sony | Accountant | 1979 | UK | Married |
| 4013 | SeolDiao | Male | Chester University | Master's | Maxtor | Database Administrator | 1983 | France | Single |
| 4014 | Lore Parsan | Female | Pietersburg University | Bachelor | Sonic | Database Administrator | 1982 | Spain | Single |
| 4023 | Carolin Wolf | Female | Franklin High school | Diploma | Sony | Bookkeeping | 1979 | Germany | Married |
| 1081 | Alex Monata | Male | Lowa University | Master's | Sony | Electrical Engineer | 1986 | UK | Married |

| 37 | Silvia Jacson | Female | Carolina University | Bachelor | MySpace | Computer Data Clerk | 1978 | Australia | Married |
| --- | --- | --- | --- | --- | --- | --- | --- | --- | --- |
| 1187 | Shery Monaten | Female | Dublin High school | Diploma | MySpace | Buyer | 1968 | Australia | Single |
| 1195 | Melina Diyana | Female | Pietersburg University | PhD | MySpace | Call Center Assistant | 1989 | Australia | Single |
| 1234 | LinaEghose | Female | Gabelino High school | Diploma | Amgen | Buyer | 1980 | Canada | Single |
| 1235 | MariyanaPlanta | Female | Iowa University | Bachelor | Amgen | Electrical Engineer | 1987 | Canada | Single |
| 1237 | Toney Cazola | Male | Carolina University | Bachelor | Amgen | Call center Operator | 1978 | Canada | Single |

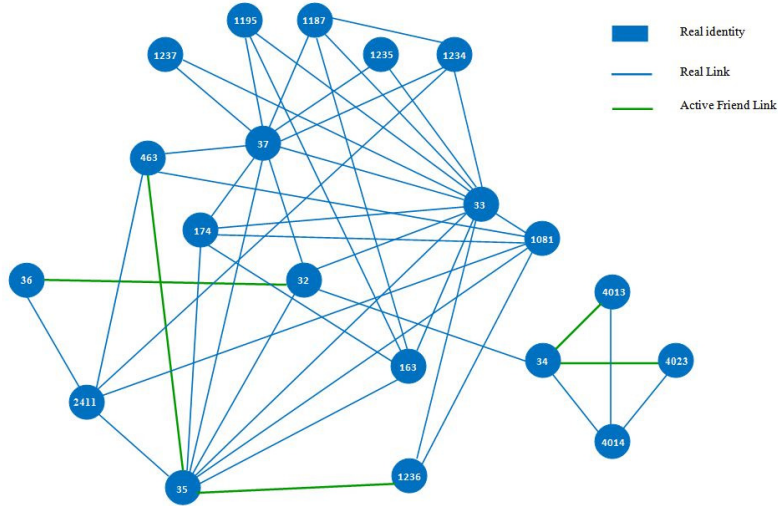

Figure 7.20 users' social graph

As mentioned before, an active friend is a friend who posts on the wall, comments and tags on her/his friends' posts. This relationship is shown in green lines in Figure7.

## 4.1. Testing the IAC approach on dataset

All of detection steps (6 steps) are applied to users of dataset as well as it is supposed that they do not use any particular privacy setting.

### 4.1.1. Choosing a victim identity

Initially, a user is selected as a victim identity from dataset. As it is mentioned in section 1, an attacker makes a fake identity considering some acceptable information of a real identity which he/she has already gathered from online social networks or other sites. Attacker uses this victim to reach his goal by connecting to victim's friends [4].User 35 is chosen as a victim because it has some perquisites as the number of links (edges) and social activities (green edges) in the network. Therefore a victim identity 35′ is created and its attribute values are displayed in Table 2 and Figure 8 demonstrates its position in social graph in red color.

Table 2. 20 users' attributes selected from dataset with fake identity

| ID | Name | Gender | Education | | Work | | Birthday | Location | Relationship |
|---|---|---|---|---|---|---|---|---|---|
| | | | School | Degree | Employer | Position | | | |
| 32 | NikoParda | Female | Harvard University | PhD | East Man | Manager | 1979 | USA | Single |
| 35 | Sara Abraham | Female | Arcadia University | Master's | Owens | Web Developer | 1980 | USA | Single |
| 35′ | Sara Abraham | Female | Arcadia University | Bachelor | Owens | Web Developer | 1980 | USA | Single |
| 36 | Sara Abraha | Female | Carolina University | Master's | Owens | Web Developer | 1980 | USA | Single |
| 174 | David Ernox | Male | Michigan University | Master's | Qpass | Java Developer | 1984 | USA | Single |
| 463 | Sara Abram | Female | Michigan University | Master's | AppNet | Web Developer | 1985 | USA | Single |
| 1236 | Tom Banho | Male | Acaedia University | Bachelor | Xing | Network Manager | 1979 | USA | Married |
| 2411 | Rose Milan | Female | Koln University | PhD | Axvert | Manager | 1972 | USA | Single |
| 33 | HanrryDabuo | Male | Dublin High school | Diploma | Sonic | Secretary | 1970 | UK | Married |
| 34 | Rosa Morada | Female | Franklin High school | Diploma | Sonic | Bookkeeping | 1974 | UK | Married |
| 163 | CharlsSelvin | Male | Pietersburg University | Bachelor | Sony | Accountant | 1979 | UK | Married |
| 4013 | SeolDiao | Male | Chester University | Master's | Maxtor | Database Administrator | 1983 | France | Single |
| 4014 | Lore Parsan | Female | Pietersburg University | Bachelor | Sonic | Database Administrator | 1982 | Spain | Single |
| 4023 | Carolin Wolf | Female | Franklin High school | Diploma | Sony | Bookkeeping | 1979 | Germany | Married |
| 1081 | Alex Monata | Male | Lowa University | Master's | Sony | Electrical Engineer | 1986 | UK | Married |
| 37 | Silvia Jacson | Female | Carolina University | Bachelor | MySpace | Computer Data Clerk | 1978 | Australia | Married |
| 1187 | SheryMonaten | Female | Dublin High school | Diploma | MySpace | Buyer | 1968 | Australia | Single |
| 1195 | Melina Diyana | Female | Pietersburg University | PhD | MySpace | Call Center Assistant | 1989 | Australia | Single |
| 1234 | LinaEghose | Female | Gabelino High school | Diploma | Amgen | Buyer | 1980 | Canada | Single |
| 1235 | MariyanaPlanta | Female | Iowa University | Bachelor | Amgen | Electrical Engineer | 1987 | Canada | Single |
| 1237 | Toney Cazola | Male | Carolina University | Bachelor | Amgen | Call center Operator | 1978 | Canada | Single |

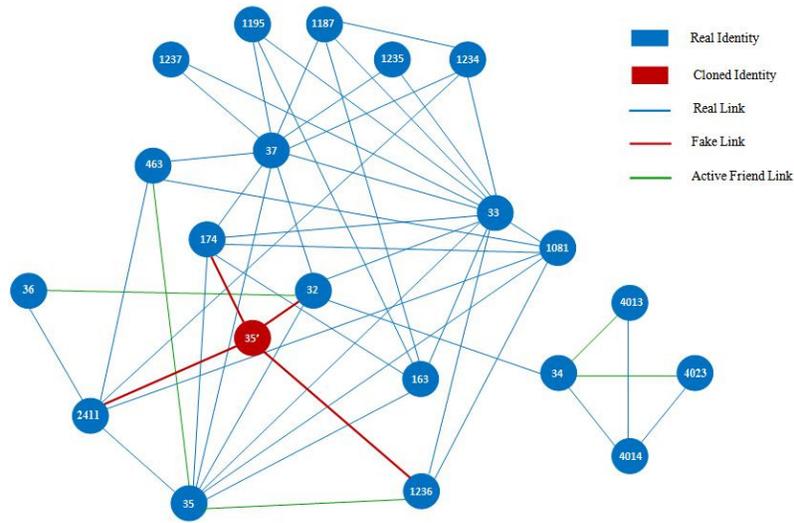

Figure 8.  20 users' social graph with a fake identity

### 4.1.2. Initializing $\alpha$

As mentioned in section 3-1, it is necessary to initialize $\alpha$ before performing the experiments on thedataset. Attribute augmented edges are chosen among the top K similar pairs of matrix C where K = $\alpha$ × |E|.  The higher mount $\alpha$ is gotten, the more edges are added to each community thus more accurate clusters are formed on the social graph. At the beginning, $\alpha$ is set by 0.68 (K = 34) then it will be set by other values in section 4-2.

### 4.1.3. Discovering communities in social graph

After performing IAC algorithm on dataset, the attribute augmented graph and clustered graph with three communities $C_1$, $C_2$ and $C_3$are gained so that isshown in Figure 9 and Figure 10 respectively.

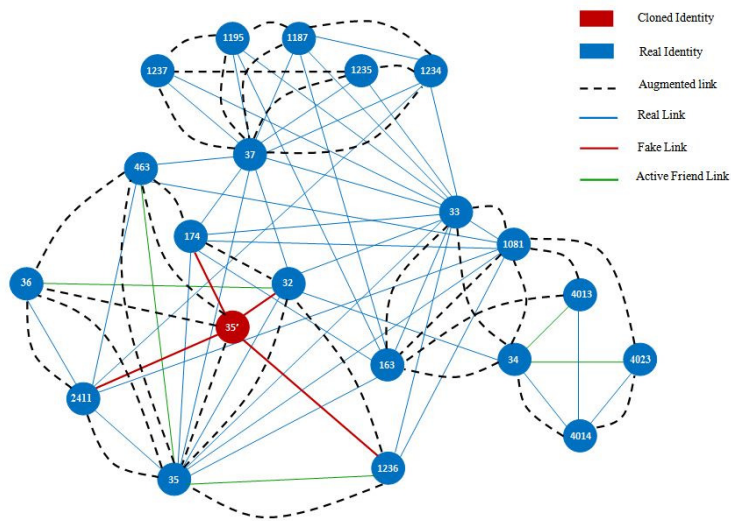

Figure 9. Attribute augmented graph

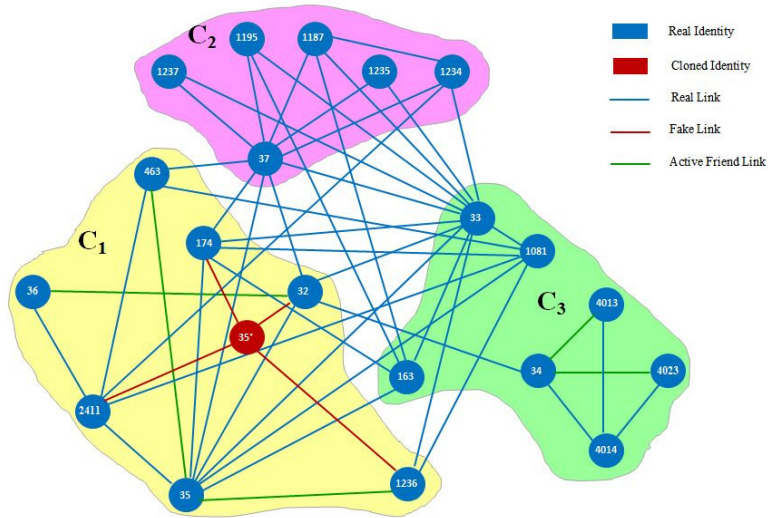

Figure 10.Clustered graph by IAC algorithm

### 4.1.4. Extracting victim's attributes

The information of victim (who wants to detect his clones) is extracted in this step and it shown in Table 3.

Table 3. Real user's attributes

| ID | Name | Gender | Education | | Work | | Birthday | Location | Relationship |
|---|---|---|---|---|---|---|---|---|---|
| | | | School | Degree | Employer | Position | | | |
| 35 | Sara Abraham | Female | Arcadia University | Master's | Owens | Web Developer | 1980 | USA | Single |

### 4.1.5. Searching in Community

Since node 35 is belong to $C_1$, only in this community is search for finding similar profiles to 35. The searchresult is shown in Table 4.

Table 4.Similar users to 35

| ID | Name | Gender | Education | | Work | | Birthday | Location | Relationship |
|---|---|---|---|---|---|---|---|---|---|
| | | | School | Degree | Employer | Position | | | |
| 35′ | Sara Abraham | Female | Arcadia University | Bachelor | Owens | Web Developer | 1980 | USA | Single |
| 36 | Sara Abraha | Female | Carolina University | Master's | Owens | Web Developer | 1980 | USA | Single |
| 463 | Sara Abram | Female | Michigan University | Master's | AppNet | Web Developer | 1985 | USA | Single |

**4.1.6. Selecting apt identities**

According to profile cloning attacks, an attacker aims victim's friends and sends them friend requests hence a cloned profile will have some victim's friends in its friend list [29]. Node 463 is not a clone identity because it is connected to node 35 directly and only 36 and 35′ are passed to next step.

**4.1.7. Computing strength of relationship**

In this step, SR is calculated for node 35′ and node 36 in regard to Eq. 2,3,4,5,6,7 then they will be ordered by values:

SR (35, 35′) = 14.497
SR (35, 36) = 36.85

As it is shown amount of SR (35, 35′) is less than other and first it will be sent to next stage for verification.

**4.1.7. Verification**

In this part, nodes 2411, 32, 1236, 174 (mutual friends between 35 and 35′) are asked to design some technical questions concerning the relationship background. Node 35 cannot answer the questions due to lack of knowledge about users pervious activities and it is marked as clone nodes.

**4.2. The role of $\alpha$ to constructing communities**

In this section, the $\alpha$ is set by some other values as represent in Table 4. For example when $\alpha=1$, the number of augmented edges will be |E|. The clustered graphs with changing $\alpha$ areshown in Figure11. If $\alpha$ is increased and the form of clustering does not change, it means that the default value for $\alpha$ was correct and most similar users are in each community.

Table 5. Different values of $\alpha$

| $\alpha$ | K |
|---|---|
| 0.68 | 34 |
| 0.78 | 39 |
| 0.88 | 44 |
| 1 | 50 |

International Journal of Network Security & Its Applications (IJNSA), Vol.6, No.1, January 2014

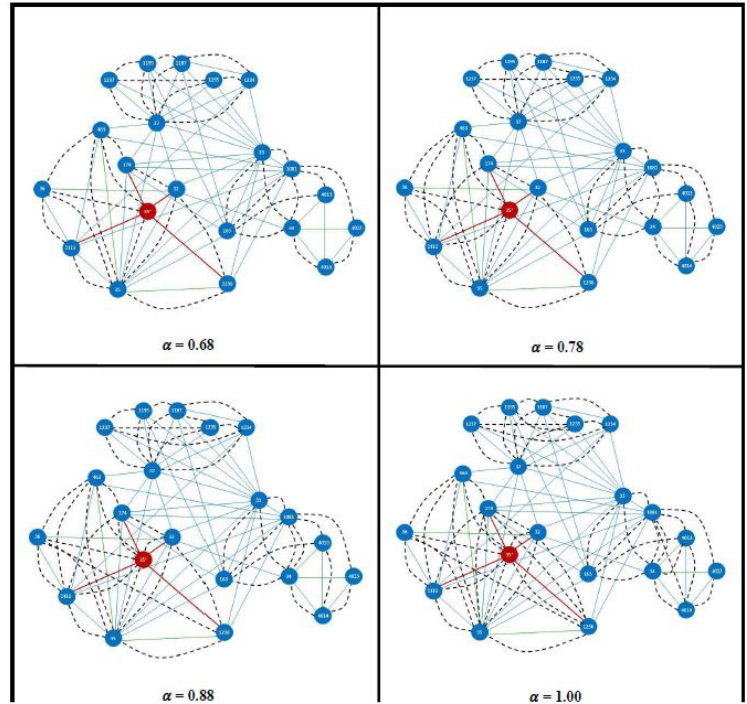

Figure 11. The different graphs with different $\alpha$

For indicating the role of $\alpha$ to construct communities with similar members, a similarity rate in cluster parameter is defined as follows:

$$Similarity\ rate\ in\ cluster = \frac{Number\ of\ agumented\ edges\ in\ cluster}{Number\ of\ edges\ in\ cluster}$$

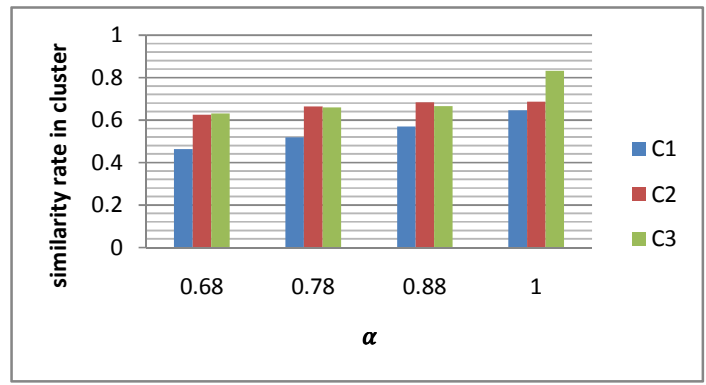

Figure 12. Similarity rate in community

The similar rate in clustersfor $C_1$, $C_2$ and $C_3$ in Figure 10, is indicated in diagram of Figure 12. According to diagram, through increasing the value of $\alpha$ the most accurate clusters are obtained in the light of similar members.

## 5. Evaluation

In order to demonstrate the accuracy of IAC approach, first two parameters are defined as follows:

True positive (TP): Number of clone nodes that are identified as fake nodes
False Positive (FP): Number of real nodes that are identified as fake nodes

Next, some other clone nodes are added to dataset and IAC approach is applied on. As shown in Figure 13, for all numbers of fake nodes, the mount of TP is higher than FP.

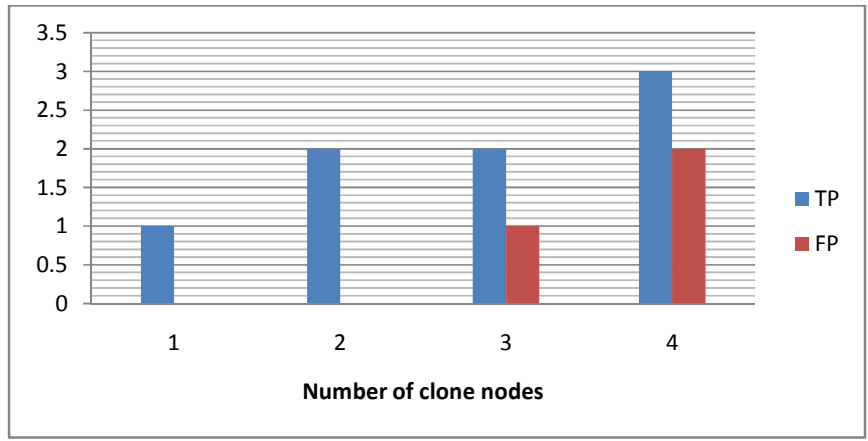

Figure 13. TP and FP for clone node detection

With the intension of comparing IAC approach to previous approaches, all of three previous approaches are applied on the dataset. As diagram in Figure 14shows, in previous approaches the mount of their TP is less than the TP of IAC approach and also the mount of their FP is more than the FP of IAC approach. Hence our approach can detect fake nodes more accurate than others.

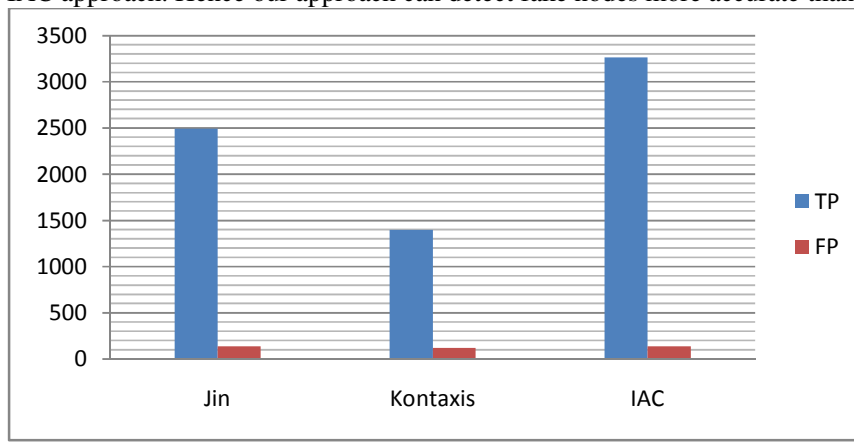

Figure 14.Comparing three exiting approaches

## 6.CONCLUSIONS

Newly, social networks became a significant part of people normal life and the most internet users spend their times on. Alongside many useful applications they have some other aspects which are growing by hackers, hustlers and online thief. In this paper, an approach was suggestedfor detecting cloned profiles depending on users' similarities and their relationship in 6 steps. It should be noted that, although detecting fake identities can stop greater extent of deception in future, prevention is better than cure because it is enough for an attacker to observer users' detail once. Therefore, teaching users is a worthy attempt to prevent cloning attacksso that they must not accept friend requests when they do not know the sender. With a view to extend the proposed approach, it can be developed as a Facebook application which each user can run it on his/her profile and also some fuzzy methods can be used to overcome wrongly typedinformation in users profiles.